\documentclass[aip,jmp,amsmath,amssymb,manuscript]{revtex4}
\usepackage{graphicx}
\usepackage{bm}
\usepackage{}

\begin{document}


\title[Manuscript draft]{Plasmons in single- and double-component helical liquids: 
Application to two-dimensional topological insulators}

\author{O. Roslyak, Godfrey Gumbs}

\affiliation{Department of Physics and Astronomy,
Hunter College, City University of New York,\\
695 Park Avenue, New York, NY 10065, USA}
\email{avroslyak@gmail.com}

\author{Danhong Huang}
\affiliation{Air Force Research Laboratory, Space Vehicles
Directorate,\\
Kirtland Air Force Base, NM 87117, USA}

%

\date{\today}

\begin{abstract}
The plasmon excitations in proposed single- and double-component helical liquid (HL) 
models are investigated within the random-phase approximation, by calculating  the
density-density, spin-density and spin-spin waves. The effect due to  broken time-reversal 
symmetry on  intraband-plasmon dispersion relation in the single-component HL system 
is analyzed and compared to those of well-known cases, such as conventional 
quasi-one-dimensional electron gases and armchair graphene nanoribbons.
The equivalence between the density-density wave in the single-component HL to 
the coupled spin-density and density-density waves in the double-component HL is
 shown here  and explained, in addition to the difference between intraband  
 and interband-plasmon excitations in these two systems. Since the two-component 
 HL can physically be thought of as a Kramers pair in two-dimensional topological 
 insulators, our proposed single-component HL model with broken time-reversal 
 symmetry, which is an artificial construct, can be viewed as an ``effective'' 
 model in this sense and its prediction may be verified in realistic systems
in future experiments.
\end{abstract}

\maketitle


\section{Introduction}
\label{SEC:1}

Topological insulators (TIs) are found to be a new class of materials
which possess insulating (large bandgap) states in the bulk but
conducting edge states on their surfaces. Only states along
edges of TIs have non-flattened dispersion (extended states) within
a bulk bandgap, thereby allowing for charge and spin currents
under zero bias\,\cite{qi2011topological}. The bulk bandgap of a
TI is usually large enough to be comparable to room temperature,
which makes TIs thermally stable and a good candidate for high-power
electrical and optical applications. One example of such TIs is the
so-called three-dimensional (3D) TI, e.g., $\texttt{Bi}_2 \texttt{Te}_3$
crystals. These 3DTIs are 3D band insulators possessing two-dimensional
(2D) conducting surface states. Additionally, there exist 2DTIs, in
which the conducting states are localized close to all edges of a
slab in real space and display a well-defined Dirac cone in momentum
space. In this paper, we confine our attention to another predicted
type of TI which has already been demonstrated in an inverted
$\texttt{HgTe/CdTe}$ (also called type-III) quantum well
when the well thickness exceeds $6.3$\,nm. The type-III quantum
well is formed by sandwiching a thin narrow-bandgap $\texttt{HgTe}$
layer between two thick wide-bandgap $\texttt{CdTe}$ layers in the
growth ($z$) direction. Bulk $\texttt{HgTe/CdTe}$ is a semimetal
with zero bandgap. The quantum-size effect from a quantum well
introduces a finite but very small bandgap to a $\texttt{HgTe/CdTe}$
layer. As expected, the induced bandgap in a $\texttt{HgTe/CdTe}$ layer
decreases with the layer thickness. For an inverted $\texttt{HgTe/CdTe}$
quantum well, the usual lower $p$-type $\Gamma_8$-band moves above
the $s$-type $\Gamma_6$-band around the center of the Brillouin zone
to create an anti-crossing bulk bandgap as well as a negative effective
mass for conduction electrons at the same time. This $\texttt{HgTe/CdTe}$-based
TI is one type of 2DTI. A semi-infinite quantum well extends infinitely
in the $x$ direction but is still confined within a half-plane ($y<0$)
 in the $y$ direction. Consequently, there always exists one conducting
 channel (helical edge state) for spin-up or spin-down electrons along
 the edge ($y=0$) of a half-$xy$ plane (with a finite thickness in the
 $z$ direction). This means that we may easily assign a spin component
 to a current if we know its flowing direction. The switching from a
 left to a right circular-polarization of incident light is expected
 to change the flow direction of a photocurrent\,\cite{nature-gedik}.
This helical state spans over a quantum well in the $z$ direction and
forms a quasi-one-dimensional electron gas (quasi-1DEG). The energy
dispersion of the helical state lies within the anti-crossing bandgap
region of an inverted $\texttt{HgTe/CdTe}$ quantum well. The theoretical
description of this type of topological states is given by the model of
Bernevig, Hugues and Zhang (BHZ)\,\cite{könig2008quantum,bernevig2006quantum}.
The general classification of TI is present in Refs.\,\cite{schnyder2008,kitaev2009}.

\medskip

The spin dynamics of TIs have received a great deal of attention,
including a topological quantum phase transition in a tunable spin-orbit system\,\cite{phase}, spin-polarized electrical
current\,\cite{raghu2010collective} and a photocurrent induced by  circularly-polarized incident light\,\cite{nature-gedik}.
However, unique properties of charge dynamics in the same
systems are much less known. The goal of this paper is to investigate
the charge density/spin dynamics of collective excitations of those edge-bound
electrons.
To study the charge dynamics of undamped collective excitations
of electrons having  a given helicity, we employ a 2DTI model system\,\cite{könig2008quantum,bernevig2006quantum}.
Those readers interested in the collective response of 3DTIs may start by looking up Refs.
\,\cite{efimkin2011spin,raghu2010collective}.
3DTIs are beyond the scope of the present paper.
For the 2DTI system of class AII (QSH
edge), the BHZ model predicts the existence of
both insulating bulk and conducting $(1,\,0)$-edge states in an
inverted $\texttt{HgTe/CdTe}$ quantum well, similar (but not  identical) to
those in a metallic armchair graphene nanoribbon (ANR).
In a graphene ANR, the plasmon excitations are excited by interband
transitions only\,\cite{brey2007elementary}. Unlike those in a 2DTI system, electronic states in a graphene ANR are not localized around
the two ribbon edges. Moreover, both helical branches contribute
to plasmon excitations. This makes the plasmon dispersion in a
graphene ANR almost identical to that of a conventional 1DEG.
\medskip

In this paper we  investigated the effect of wave function localization
in a semi-infinite 2DTI system on the collective excitation of electrons.
Our calculated plasmon dispersion is compared with those in a conventional 1DEG\,\cite{sarma1996dynamical} and in a metallic graphene ANR.
Following the work by Zhang {\em et al.\/}\,\cite{PhysRevLett.96.106401},
we adopt the terminology that an n-component helical liquid (HL)
contains n-time reversal pairs of fermions. These n-component HL
states are localized on different edges of a quantum well. For a
semi-infinite quantum well, on the other hand, we may only need to consider
a single mode of the pair to calculate the Coulomb excitation of
electrons. 
Since throughout the paper we do restrain ourselves to the semi-infinite case 
we shall refer to such single edge mode as single component HL.
Since its counterpart is ignored the time reversal symmetry is broken. 
That is we have a quantum Hall state. Recall that the quantized charge Hall (QH)
conductivity is attributed to the transport via a single chiral edge mode. 
\par
On the other hand when we do include finite spin into consideration we would have quantum spin Hall (QSH) effect which does not require time reversal breaking.
A pair of states per edge appear. Here we refer to such case as two component HL thus effectively considering $Z_2$-trivial TI\cite{könig2008quantum}.
The contributions from a single
and two component HL to the plasmon excitation are explored.
Our calculations indicate that ANR and the semi-infinite 2DTI system
can be considered to be two and single component HL, respectively,
in calculating the charge dynamics of collective excitations.
\medskip

The rest of the paper is organized as follows. In Sec.\,\ref{SEC:2},
we calculated the edge-localized helical states of electrons and their
energy dispersion in a semi-infinite inverted $\texttt{HgTe/CdTe}$
quantum well. In Sec.\,\ref{SEC:3}, the dispersion of edge-plasmon
excitations of single component HL with broken time-reversal symmetry is
calculated. That type of density-density response is compared with that of a conventional one-dimensional
electron gas, a metallic armchair graphene nanoribbon and finally with the response of the Kramers pair of two-component HL.
In Sec.\, \ref{SEC:4} we argue that the density-density plasmon excitations of single component HL are formally equivalent to the interference pattern between
spin-density and density-density plasmons in two component HL
Finally, the conclusions of the paper are given briefly in Sec.\,\ref{SEC:5}.

\section{Formalism of 2DTI in $\texttt{HgTe/CdTe}$ quantum well}
\label{SEC:2}

To study the collective electronic excitations, let us start with the
BHZ model for the electron band structure near the center
$\Gamma_k=(0,\,0)$ of the first Brillouin zone. We assume that
the quantum well is infinite along the $x$ direction and
finite or semi-infinite along the $y$ direction.
The width of the well is given implicitly within the parameters
[$A$, $B$ and $\Delta$ in Eq.\,(\ref{EQ:1_1})] of the model Hamiltonian.
The ``ansatz'' wave function is taken to be $\exp(ikx)$ in the $x$
direction. Along the $y$ direction, we discretize the spatial position as
$y=ja>0$ with $j=1,\,2,\,\ldots,\,N$ being a positive integer and $a$
being the lattice constant, where $N$ is the total number of sites
assumed for numerical simulations in the $y$ direction, and is
 taken sufficiently large to ensure that two edge topological states
 do not overlap each other. The wave number $k$ in the $x$ direction
 is given in units of $k_B=\pi/a$, leading to discrete spatial positions
  along the $x$ axis as $x=\ell a$ with $\ell=-N/2,\,\ldots,\,-2,\,-1,\,0,\,1,\,2,\,\ldots,\,N/2$
  being an integer.
\medskip

According to  Ref.\,\onlinecite{könig2008quantum}, the Hamiltonian
describing electronic states in a $\texttt{HgTe/CdTe}$ (type-III)
quantum well can be written as a block-tridiagonal matrix

\begin{align}
\label{EQ:1_1}
&\mathcal{H}
=
\left({
\begin{array}{ccccccc}
\ddots & \ddots & \ddots & 0 & 0 & 0 & 0\\
0 & t & \varepsilon & t^\dag & 0 & 0 & 0\\
0 & 0 & t & \varepsilon & t^\dag & 0 & 0\\
0 & 0 & 0 & t & \varepsilon & t^\dag & 0\\
0 & 0 & 0 & 0 & \ddots & \ddots & \ddots \\
\end{array}
}\right)_{4N \times 4N}\ ,\\
\notag
&\varepsilon = A\,\texttt{sin}(k)\,\Gamma_1+\left[\Delta-4B+2B\,\texttt{cos}(k)\right]\,\Gamma_5\ ,\\
\notag
&t=-\frac{i A}{2}\,\Gamma_2+B\Gamma_5\ .
\end{align}
Here, the elements of the Clifford algebra are expressed
in terms of $\Gamma_1=\sigma_x\otimes\sigma_z$,
$\Gamma_2=-\sigma_y\otimes\sigma_0$ and
$\Gamma_5=\sigma_z\otimes\sigma_0$
with $\sigma_i$ denoting the Pauli matrices. Additionally,
$A,\,B,\,\Delta$ in Eq.\,(\ref{EQ:1_1}), which are scaled
by $\hbar k_Bv_F$ with $v_F\sim c/1000$ being the Fermi velocity
expressed in terms of the speed of light $c$, are the
material parameters dependent on the quantum well width.
The form of the Hamiltonian in Eq.\,\eqref{EQ:1_1} implies
that the associated wave function vanishes at edges.
\medskip

\begin{figure}[]
\centering
\includegraphics[width=0.5\textwidth]{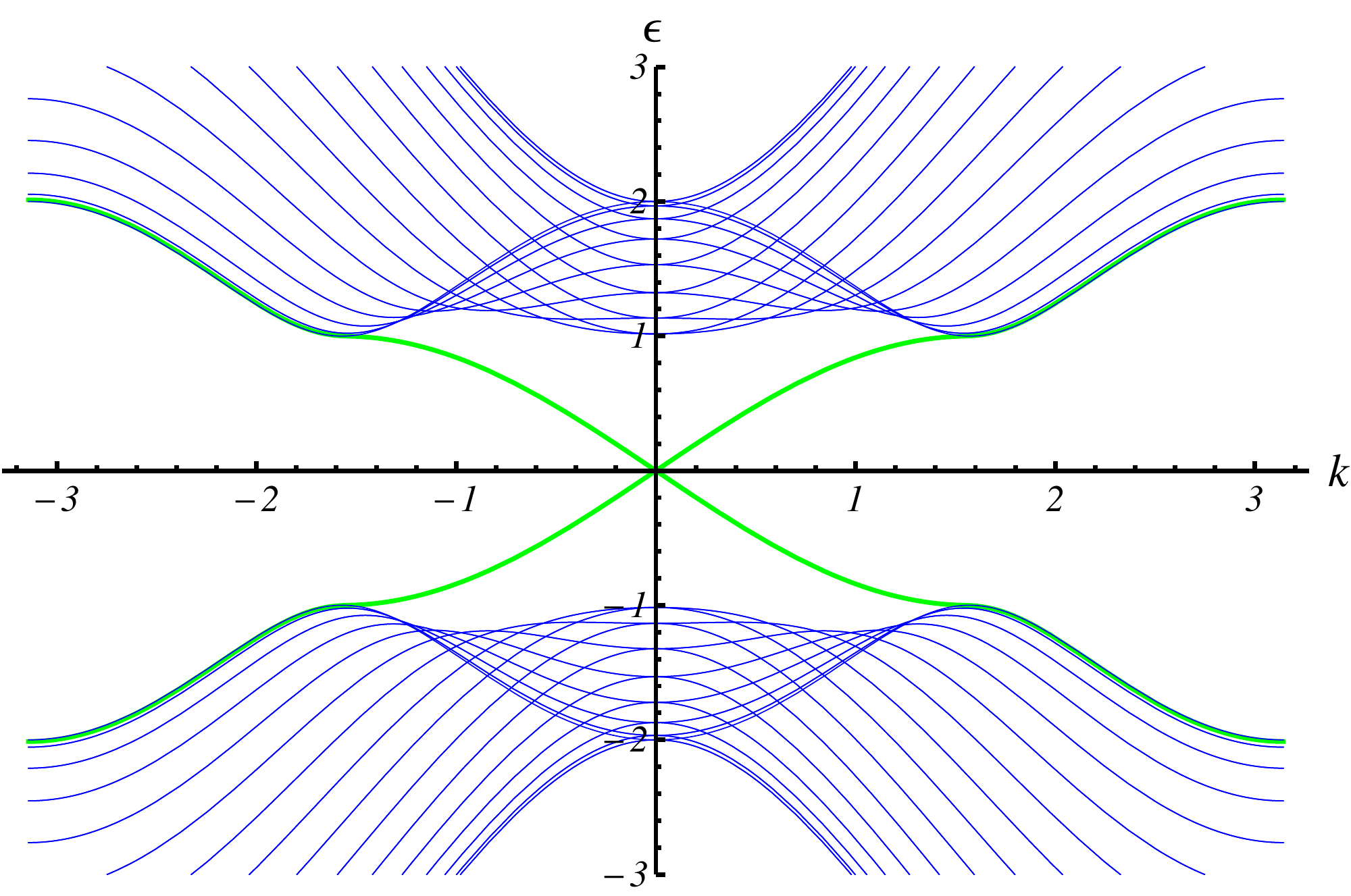}
\caption{\label{FIG:1} (Color online) Full band structure
$\varepsilon(k)$ of a 2DTI with $\Delta=2B$, $A=B=1$ and
$N=30$. The blue curves correspond to the bulk modes while
the two green curves represent the topological surface states.
Each state is doubly degenerate with respect to the two
eigenvalues of the $\sigma_x$ matrix.}
\end{figure}

The calculated energy dispersion corresponding to Eq.\,(\ref{EQ:1_1})
is presented in Fig.\,\ref{FIG:1}. For $0<\Delta/B<4$, two localized
states at the boundaries are obtained, which may be expressed in terms
of  a linear combination of the ansatz wave functions\,\cite{imura2010zigzag}

\begin{gather}
\label{EQ:1_2}
\Psi_{j,\,l}=\frac{1}{\sqrt{Na}}\sum\limits_k\,\psi^>_j(k)\,
\texttt{e}^{ikl}+\frac{1}{\sqrt{Na}}\sum\limits_k\,
\psi^<_j(k)\,\texttt{e}^{-ikl}\\
\notag
=\frac{1}{\sqrt{Na}}\sum \limits_k\,\left[\rho(k)\right]^j\,
\texttt{e}^{ikl}  +  \frac{1}{\sqrt{Na}}\sum\limits_k\,\left[\rho(k)\right]^{N+1-j}\,\texttt{e}^{-ikl}\ ,
\end{gather}
where the helical states are chosen to be the eigenstates defined by $\Gamma_1\vert{\pm}\rangle=\pm\,\vert{\pm}\rangle$.
The analytic solutions of Eq.\,(\ref{EQ:1_2}) may be obtained
by substituting the above ansatz wave functions into
Eq.\,\eqref{EQ:1_1} and considering the identity $\left[i\Gamma_5\Gamma_2,\,\Gamma_1\right]=0$. This yields

\begin{align}
\label{EQ:1_3}
&\psi^>_j(k) =
\left({
c^>_{+,1}[\rho_{1}(k)]^j + c^>_{+,2}[\rho_{2}(k)]^j
}\right) \vert{+}\rangle\\
\notag
&+ \left({
c^>_{-,1}[\rho_{1}(k)]^{-j}+ c^>_{-,2}[\rho_{2}(k)]^{-j}
}\right) \vert{-}\rangle\ ,\\
\notag
&\psi^<_j(k) =
\left({
c^<_{-,1}[\rho_{1}(-k)]^{N+1-j}+ c^<_{-,2}[\rho_{2}(-k)]^{N+1-j}
}\right) \vert{-}\rangle\\
\notag
&+ \left({
c^<_{+,1}[\rho_{1}(-k)]^{-N-1+j}+ c^<_{+,2}[\rho_{2}(-k)]^{-N-1+j}
}\right) \vert{+}\rangle \ .
\end{align}

\medskip

\begin{figure}[]
\centering
\includegraphics[width=0.5\textwidth]{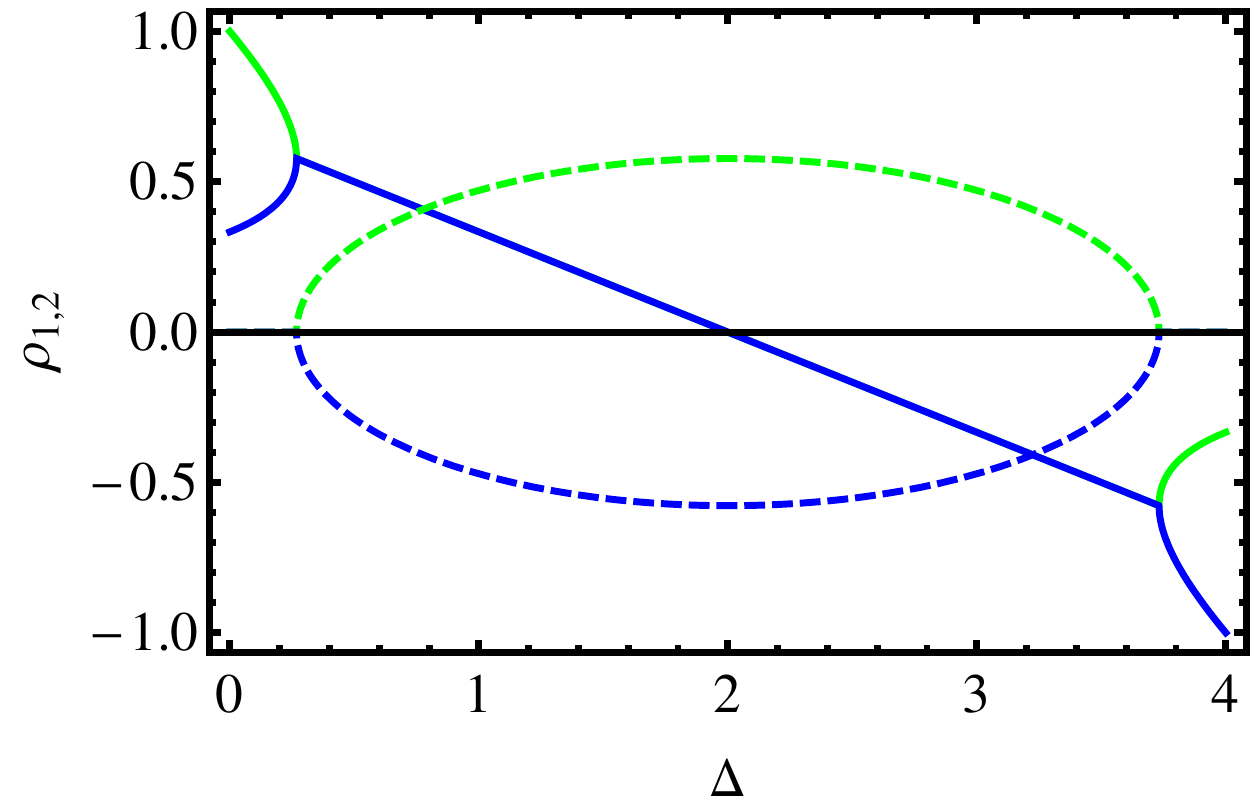}
\caption{\label{FIG:4} (Color online) Based on Eq.\,(\ref{rho1+2})
the real (solid curves) and imaginary (dashed curves) parts of
$\rho_{1}$ (green) and $\rho_{2}$ (blue) at $k=0$ as a function of $\Delta/B$ for $A=B=1$.}
\end{figure}

The wave function corresponding to $\vert {+} \rangle$ in
Eq.\,\eqref{EQ:1_3} is related to the energy dispersion
$E_+(k)=A\,\texttt{sin}(k)$, while that corresponding to $\vert {-} \rangle$ is associated with $E_-(k)= -A\,\texttt{sin}(k)$,
as shown by two green curves in Fig.\,\ref{FIG:1}. The deviation from
the sine function becomes significant once $E_{\pm}(k)$ merges
with the bulk modes (shown as blue curves). Parameters $\rho_1(k)$
and $\rho_2(k)$ in Eq.\,(\ref{EQ:1_3}) are defined by

\begin{gather}
\rho_{1,\,2}(k) = \frac{-\Omega(k) \pm
\sqrt{\Omega^2(k) + A^2 - 4 B^2}}{A + 2 B}\ ,\\
\Omega(k) = \Delta -2B\left[{2-\texttt{cos}(k)}\right]\ .
\label{rho1+2}
\end{gather}
For $A=B=1$ and $k=0$, the variation of these parameters with respect to $\Delta/B$ is displayed in Fig.\,\ref{FIG:4}.
\medskip

Since our interest is limited to calculating the plasmon dispersion
in the long-wavelength limit, we would only consider the case with
$k\ll 1$. This leads to the approximate expression
$E_{\pm}(k)\approx\pm Ak$, and $\rho_{1,\,2}(k)\approx\rho_{1,\,2}(0)$
become $k$-independent at the same time.
For $\Delta = 2B$, we find $\rho_{1,\,2} = \pm i/\sqrt{3}$.
\medskip

The coefficients $c_{\pm,\,\alpha}$ with $\alpha=1,\,2$ in
Eq.\,\eqref{EQ:1_3} may be determined from the boundary conditions
as well as the wave function normalization condition. By choosing
$N\gg 1$, the condition for a finite-valued wave functions
requires $c^>_{-,\,\alpha}=c^<_{+,\,\alpha} =0$. Moreover,
the vanishing boundary conditions lead to $c^>_{+,\,1}+c^>_{+,\,2}=0$ and $c^<_{-,\,1}+c^<_{-,\,2}=0$.
After normalizing the wave functions in Eq.\,\eqref{EQ:1_3}, we
find $c^>_{+,\,1} = c^<_{-,\,1} = 2/\sqrt{3}$. Therefore, the two components of the wave
function in Eq.\,\eqref{EQ:1_2} do not overlap and are localized on opposite boundaries.
In addition, the eigenvalues of the spin state of $\sigma_x$ are
 related to the sign of the group velocity,
i.e., $\psi^>_j$ corresponds to $E_{+}(k)$ while $\psi^<_j$ is
related to $E_{-}(k)$. This pair of fermions constitutes a
$1$-component HL, and is connected by the time-reversal symmetry.
For a semi-infinite quantum well with $N \rightarrow \infty$ and
$\rho_{1,\,2} = \pm i/\sqrt{3}$, on the other hand,
we are left with a single helical state localized at the
$j=0$ boundary

\begin{gather}
\label{EQ:1_4}
\psi^>_j \approx \frac{2(i)^j}{(\sqrt{3})^{j+1}}\left[{1-(-1)^j}\right] \vert{+}\rangle\ ,\\
\notag
E_+(k)\approx Ak\ .
\end{gather}
If the material parameters $\vert{\rho_{1,\,2}}\vert>1$
 are chosen, one would retain the left mover $\psi^<_j$
 proportional to $\vert {-} \rangle$ as the proper solution.
For a $2$-component HL (Kramers pair), we must consider two helical states
on each edge. Those two states are related in Eq.\,\eqref{EQ:1_3} by the spin change $\vert+\rangle \leftrightarrow \vert - \rangle$ and the time reversal $k \leftrightarrow - k$ as well.
Therefore, the second part of the Kramers pair is:
\begin{gather}
\label{EQ:1_4_1}
\psi^>_j \approx \frac{2(i)^j}{(\sqrt{3})^{j+1}}\left[{1-(-1)^j}\right] \vert{-}\rangle\ ,\\
\notag
E_-(k)\approx - Ak\ .
\end{gather}

\section{Plasmons in HL compared with 1DEG}
\label{SEC:3}

Based on the calculated full band structure from the BHZ model, we
will further study the electron screening dynamics from the
dielectric function of a 2DTI system.
The edge states in such a system appear as the Kramers pair (Eq.\,\eqref{EQ:1_4})
with the electron spin attached to its momentum.
On the  level of density-density response it is imposable to break up the Kramers pair into its components, thus
observing the response of the two component HL.
However in the next section we shall show that the density-density response of the
single component HL formally correspond to the interference pattern of the density-density and spin-density
waves of the two component HL.

\par
\begin{figure}[]
\centering
\includegraphics[width=0.5\textwidth]{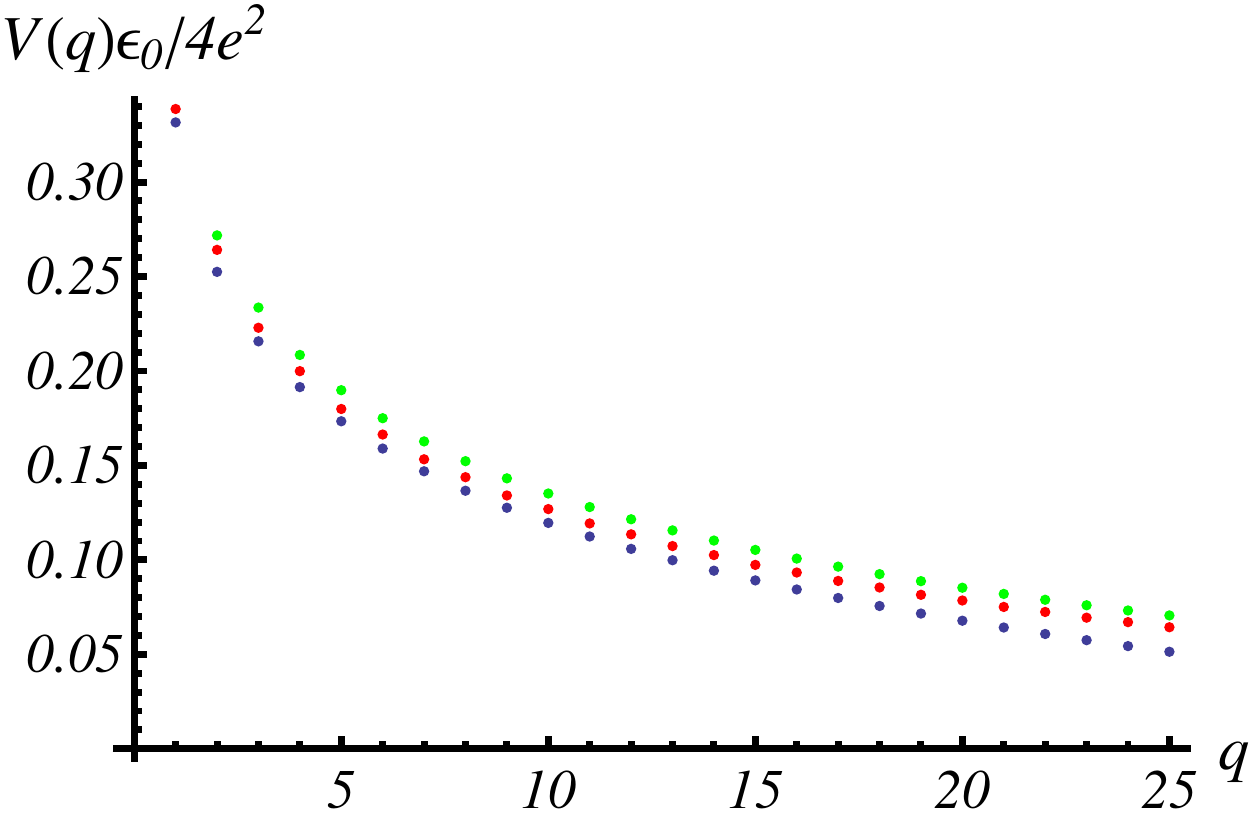}
\caption{\label{FIG:2} (Color online)
The $q$ dependence (in units of the reciprocal inter-atomic spacing $a$)
of the scaled Coulomb matrix element $V(q)/(4e^2/\epsilon_sa)$
 in Eq.\,\eqref{EQ:2_4} evaluated   between its upper
 $(3/40)\,K_0(q)$ (green) and lower bound $-(3/40)\,\ln(q)$
 (black) for various values of the ratio $\Delta/B$, where $A=B=1$.
The result for the approximate expression in Eq.\,(\ref{EQ:2_4_1}) is given by red dots in the figure.}
\end{figure}

We shall formally start with the density-density response of the single component HL.
Specifically, we will consider
a semi-infinite type-III quantum well, in which only one  helical
state can occur and is localized around the edge ($y=0$) of the system.
As a result, there exist only intraband transition in our system.
For the wave function given by Eq.\,(\ref{EQ:1_4}), the plasmon
excitation dispersion $\omega_p(q)$ within the random-phase
approximation (RPA) is determined by the
zero of the following dielectric function

\begin{gather}
\label{EQ:2_1}
\epsilon(q,\,\omega) = 1 - V(q)\,\Pi_{+,+} (q,\,\omega+i0^+)\\
\notag
=1-\frac{1}{\pi}\,\frac{q V(q)}{\hbar\omega - A q} + i  q\,V(q)\,\delta(\hbar\omega - A q) \ ,
\end{gather}
where the  noninteracting polarization function at zero temperature
is given by

\begin{gather}
\label{EQ:2_2}
\Pi_{+,+}(q,\,\omega+i0^+) = \frac{1}{N} \sum \limits_k\,
\frac{\theta(k+q)-\theta(k)}{E_+(k) - E_+(k+q) + \hbar\omega + i0^+}\\
\notag
= \frac{2}{2\pi}\,\frac{q}{\hbar\omega - A q + i0^+} \ .
\end{gather}
Here, $\theta(x)$ is the unit step function and $E_F=0$ is
assumed for the Fermi energy. The perfector of two in
Eq.\,(\ref{EQ:2_2}) accounts for the double degeneracy of the
bands. Very importantly, the  time-reversal symmetry is
broken in Eq.\,\eqref{EQ:2_2} for the response function, i.e.,
$\Pi_{+,+}(q,\,\omega+i0^+) \neq \Pi^\ast_{+,+}(q,\,-\omega+i0^+)$.
The Coulomb matrix element introduced in Eq.\,(\ref{EQ:2_1}) is given by\,\cite{PhysRevB.43.11768}

\begin{equation}
\label{EQ:2_4}
V(q) = \frac{2 e^2}{\epsilon_sa}
\sum \limits_{j=1}^{\infty}\,\sum \limits_{j'=j+1}^{\infty}\,
\vert{\psi^>_j}\vert^2 \vert{\psi^>_{j'}}\vert^2\,
K_{0}\left[{q (j'-j) }\right]\ ,
\end{equation}
where $K_0(x)$ is the modified Bessel function of the second kind, $\epsilon_s=\epsilon_0\epsilon_b$ and $\epsilon_b$ is the dielectric
constant of the host material. If it is not explicitly stated, our
parameters of choice are $A=B=1$ and $\Delta =2B$ in this paper.
For this parameter choice, the summation in Eq.\,(\ref{EQ:2_4}) can be carried out explicitly to give

\begin{gather}
\label{EQ:2_4_1}
V(q)=
\frac{2 e^2}{\epsilon_sa}\int\limits_{0}^{\infty}\,
\frac{12 \texttt{cos}(q t)\,dt}{5\sqrt{1+t^2}\left[{41-9 \texttt{cos}(2qt)}\right]}\\
\notag
\approx
\frac{e^2}{\epsilon_sa}\,\frac{3}{20}\,K_0(q)\ .
\end{gather}
Here, the validity of the last approximation is demonstrated
in Fig.\,\ref{FIG:2}, and the perfector can be calculated, by using
the fine-structure constant and $\epsilon_b=7$ for $\texttt{HgTe}$, as
$e^2k_B/\epsilon_sA=4\pi\times 0.001(\hbar k_Bv_F/A)(c/v_F)\approx 4\pi$.
\medskip

\begin{figure}[]
\centering
\includegraphics[width=0.5\textwidth]{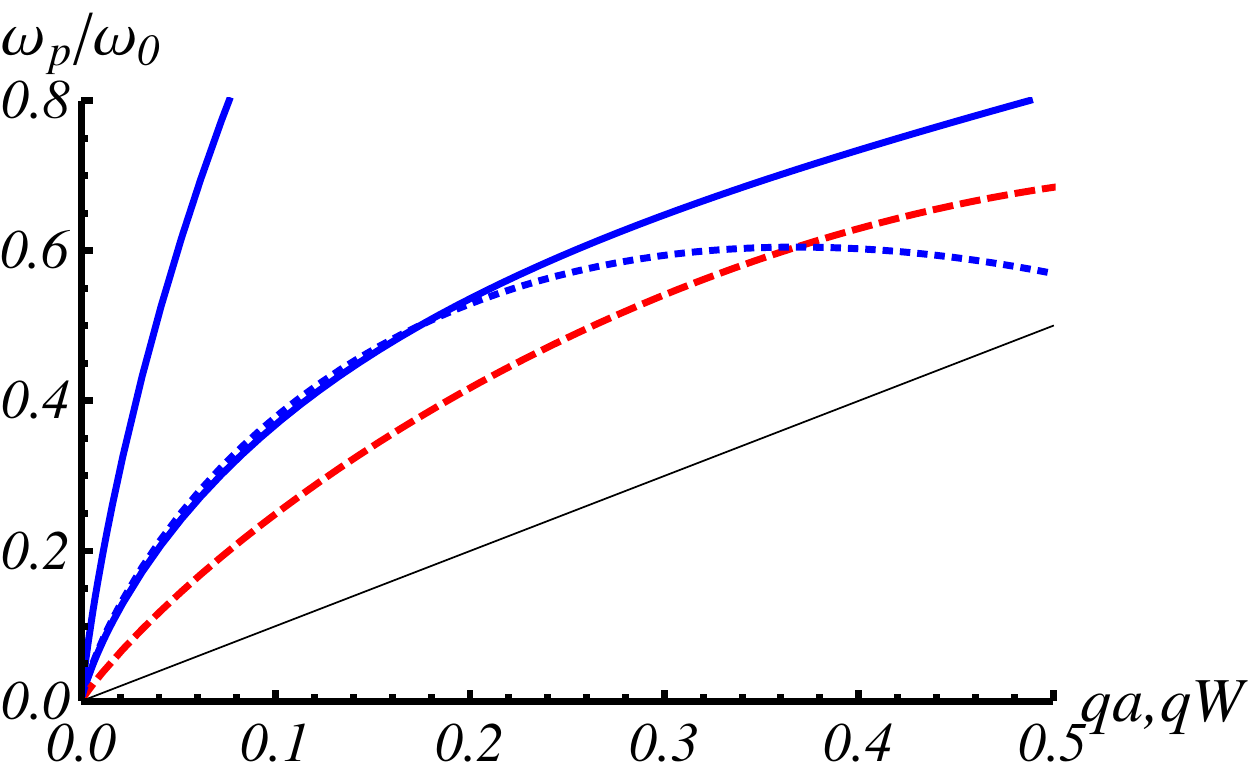}
\caption{\label{FIG:3} (Color online)
Scaled plasmon excitation energy $\omega_p(q)/\omega_0$
 of 2DTI (blue solid curve) given by Eq.\,\eqref{EQ:2_5} with
 $A=B=1$. The lower blue solid curve corresponds to $\Delta=2B$,
 while the upper blue solid curve is for $\Delta=B$ or $\Delta =3B$.
The result of Eq.\,(\ref{EQ:2_6}) is given by the  dashed blue
curve in the long-wavelength limit. The conventional 1DEG plasmon
dispersion is indicated by a red dashed curve. The particle-hole excitation region is represented by a thin black
straight line.}
\end{figure}

The imaginary part of the dielectric function in Eq.\,\eqref{EQ:2_1}
yields the boundary $\hbar\omega_{p-h} = A q$ for the particle-hole
excitation region.
Collapse of the particle-hole region into the single line is characteristic to the linear dispersion\,\cite{brey2007elementary}.
Zeros of the real part give rise to the plasmon
dispersion relation:

\begin{equation}
\label{EQ:2_5}
\hbar\omega_p(q) =  A q \left[{1 + \frac{1}{\pi A}\,V(q)}\right]\ .
\end{equation}
In the long-wavelength limit with $q a \ll 1$, we obtain $\texttt{Re}[\Pi_{+,+}(q,\,\omega+i0^+)]=q/\pi\hbar\omega + {\cal O}(q^2)$
as a leading-order result. In conjunction with Eq.\,\eqref{EQ:2_4_1},
the above equation leads to the explicit plasmon dispersion relation in the long-wavelength limit:

\begin{gather}
\label{EQ:2_6}
\omega_p(q)=-\omega_{0}\,q\ln(q) + {\cal O}(q^3)\ ,\\
\notag
\hbar\omega_0 = \frac{e^2}{\pi\epsilon_sa}\,\frac{3}{20}\ .
\end{gather}
Results from our calculations  based on Eqs.\,\eqref{EQ:2_4_1},
\eqref{EQ:2_5} and \eqref{EQ:2_6} are presented in Fig.\,\ref{FIG:3}.
It is found that $A=B=1$ and $\Delta=2B$ gives us the lower boundary
of the plasmon excitation energy. This implies that the variations
of $A$, $B$ and $\Delta$ within the regime in which the topological
edge state exists results in an increase in the plasmon energy.
\medskip

\begin{figure}[]
\centering
\includegraphics[width=0.5\textwidth]{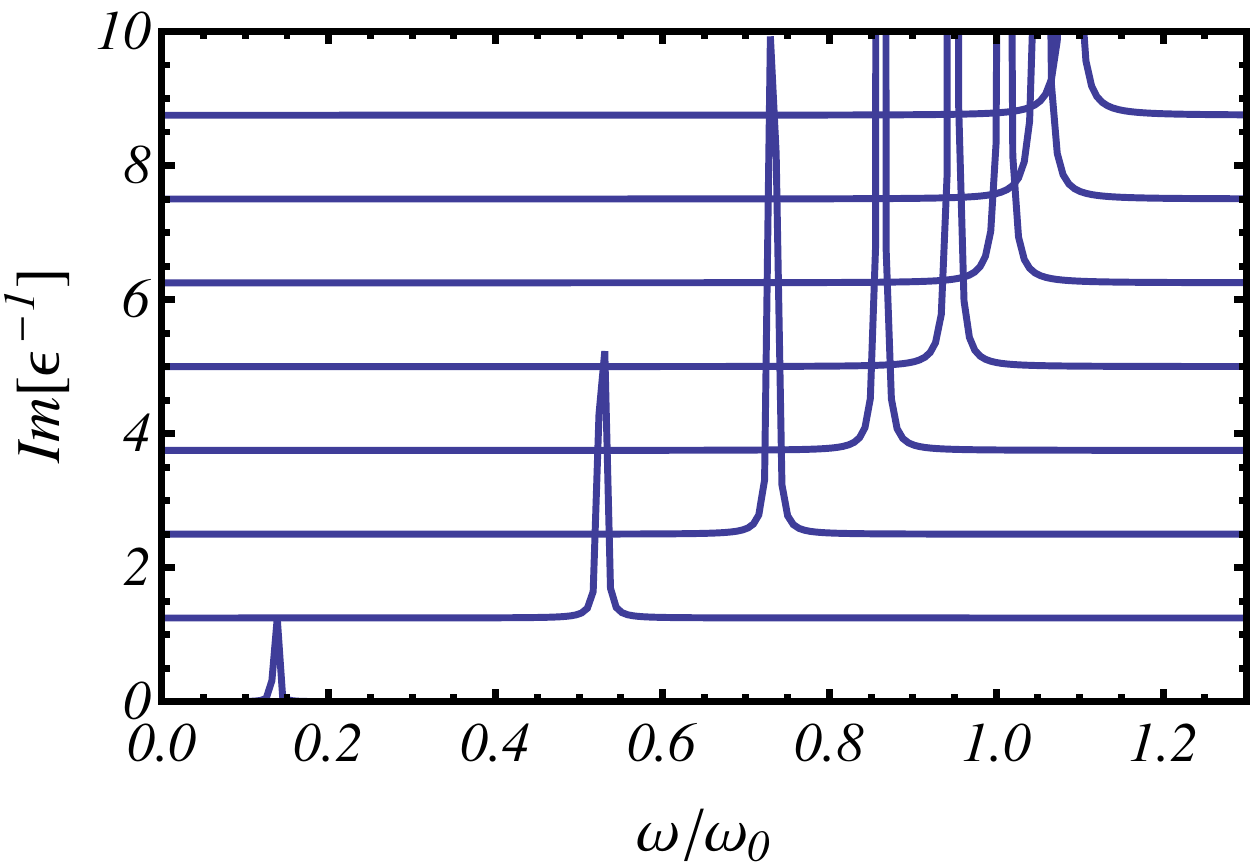}
\caption{\label{FIG:6}
Spectral function $\texttt{Im}[\epsilon^{-1}(q,\,\omega)]$ of
single component HL in 2DTI as a function of $\omega/\omega_0$ with
and chosen $q$. Here, the curves with different values for $q$
are offset vertically for clarity.
}
\end{figure}

Most  inelastic scattering experiments measure the dynamic structure
factor $\sim \texttt{Im}[\epsilon^{-1}(q,\,\omega)]$ or the inverse
dielectric function. We display $\texttt{Im}[\epsilon^{-1}(q,\,\omega)]$
in Fig.\,\ref{FIG:6} as a function of $\omega$ for chosen values of
$q$. Clearly,  the spectrum of $\texttt{Im}[\epsilon^{-1}(q,\,\omega)]$
is dominated by the plasmon resonance. The particle-hole excitation
is not pronounced in this figure. Around the plasmon resonances,
we can employ the plasmon-pole approximation  $\texttt{Im}[\epsilon^{-1}(q,\,\omega)]\sim\beta_q\,\delta(\omega- \omega_p)$,
where the plasmon weight is defined by

\begin{gather}
\label{EQ:2_6_1}
\beta_q = \frac{\pi}{\left[{\partial \texttt{Re}\,\epsilon(q,\,\omega) / \partial \omega }\right]_{\omega = \omega_p}}\\
\notag\\
= -\frac{\pi^2\hbar\omega_0^2 q\,\texttt{ln}^2(q)}{V(q)}\ .
\end{gather}
Although varying the parameters of a 2DTI system from those producing the minimal $\omega_p(q)$ can increase the plasmon energy, it reduces the plasmon weight.
\medskip

Now, let us compare our result with several known cases of 1DEG.
For a conventional semiconductor quantum wire with a parabolic
energy dispersion $E_0(k) = \hbar^2k^2/2m^\ast$ for conduction
electrons, the intraband plasmon dispersion in the long-wavelength
limit can be written  as\,\cite{PhysRevB.40.3169,PhysRevB.43.11768,sarma1996dynamical}

\begin{gather}
\label{EQ:2_7}
\omega_p (q)= \omega_{0}\,q\sqrt{- \ln(q)} + {\cal O}(q^3)\ ,\\
\notag
\omega_0=\left({\frac{2v_F e^2}{\pi\hbar\epsilon_s W^2}}\right)^{1/2}=\left({\frac{2n_{1D} e^2}{\epsilon_s m^\ast W^2}}\right)^{1/2}\ ,
\end{gather}
where $n_{1D}$ and $m^\ast$ denote the electron linear density and
effective mass, respectively. For 1DEG, the wave vector $q$ is
scaled with the characteristic size $W$ of the nanowire,
i.e., $q \rightarrow  q W$. The linear scaling of the plasmon
frequency with the square-root of electron density as well as the
high sensitivity to the wire characteristic size $\sim \sqrt{-\ln(q W)}$
are the unique properties of 1DEG in conventional semi-conducting nanowires.
These properties are in sharp contrast with our result in
Eq.\,(\ref{EQ:2_7}) for localized quasi-1DEG in a 2DTI system.
\medskip

\begin{figure}[]
\centering
\includegraphics[width=0.5\textwidth]{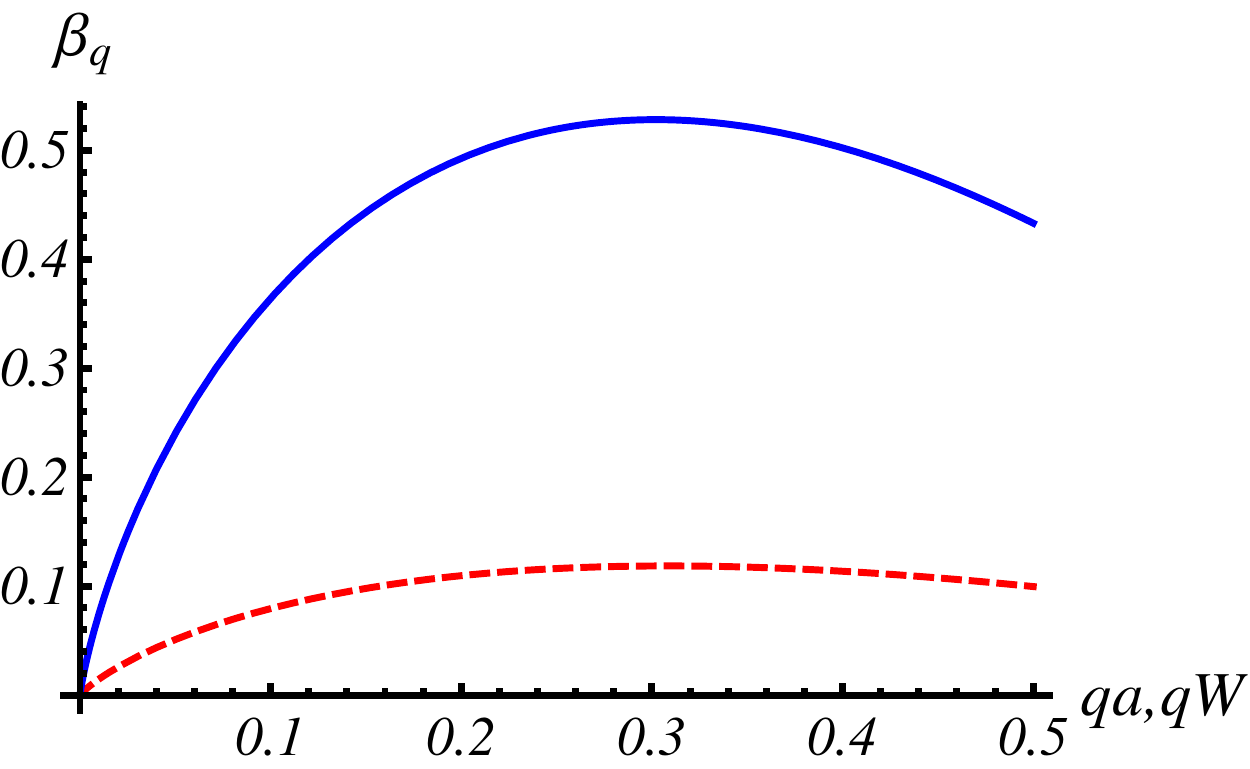}
\caption{\label{FIG:5} (Color online)
Plasmon weight $\beta_q$ for TI-based (blue solid curve) and
conventional (dashed red curve) 1DEGs, where $\omega_0$ is
 chosen as  the same for both cases.}
\end{figure}

For the conventional 1DEG, the plasmon weight is given by

\begin{gather}
\label{EQ:2_8}
\beta_q \sim  \frac{q \left[{-\texttt{ln}(q)}\right]^{3/2} }{V_c(q)}\ ,\\
\notag
V_c(q) = \frac{2 e^2}{\epsilon_sW}\left[{K_0(q) + 1.972}\right]\ ,
\end{gather}
which has different power dependence for the term $\sim\ln(q)$.
Comparison of the plasmon weights are shown in Fig.\,\ref{FIG:5}
for both the TI-based and conventional 1DEG. From Fig.\,\ref{FIG:5},
 we find that the plasmon weight in 2DTI is an order of magnitude
 larger than that of the conventional 1DEG. Consequently, a much
 more pronounced dynamical structure factor is expected for 2DTI
 [compare Fig.\,\ref{FIG:6} here with Fig.\,\,7 of Ref.\,\cite{sarma1996dynamical}]. We also note   that
 the particle-hole excitation spectrum exists in a wide region
 $k_F q/m^\ast -\hbar^{-1}E_0(q) < \omega_{p-h} <k_Fq/m^\ast + \hbar^{-1}E_0(q)$ for a conventional 1DEG rather than a narrow line in 2DTI.
\medskip

One of the striking features of 1DEG is that the RPA becomes
exact for small  value vectors  $q$. This implies that the
quasi-particles obtained from the exactly solvable linearized
Tomonaga-L\"uttinger model  agree with the calculated plasmon
dispersion relation in  the RPA\,\cite{PhysRevB.45.13713}.
In the case of a single-component HL for a 2DTI system, only
one electron branch in the Tomonaga-L\"uttinger model should
be included.
\medskip

It is also very interesting to compare the results in our paper
with similar ones for graphene-based nanostructures, where 1DEG
is provided by chiral-state electrons. A detailed calculation
using RPA for plasmon excitations in both armchair and
zigzag edged graphene-based nanoribbons was carried out  by Brey and Fertig\,\cite{brey2007elementary},
and only the armchair type was
found to exhibit undamped plasmon excitations. The most
interesting finding in Brey and Fertig's work is the presence of
metallic nanoribbons with the lowest electronic bands
given by $E_{\pm}(k)=\pm \hbar v_F \vert{k}\vert$. In these
metallic nanoribbons, the electron wave functions favor
forward scattering.
The scattering between the branches of different helicity is prohibited and corresponding structure factor vanishes.
Both helical branches (given value of the pseudo-spin) contribute equally to the response
. As a result, the polarization retains  time-reversal
symmetry. For metallic graphene nanoribbons, we obtain the interband polarization function

\begin{gather}
\label{EQ:2_9}
\chi_1(q,\,\omega+i0^+)=\Pi_{+,-}(q,\,\omega+i0^+)
+\Pi_{-,+}(q,\omega+i0^+)\\
\notag
=\frac{1}{\pi}\left[\frac{q}{\hbar\omega - \hbar v_F q + i0^+}
- \frac{q}{\hbar\omega + \hbar v_F q + i0^+}\right]\\
\notag
=\frac{2}{\pi}\,\frac{\hbar v_F q^2}{\left({\hbar\omega + i0^+}\right)^2
-(\hbar v_F q)^2}\ .
\end{gather}
Note that the polarization subindex here is related to the valley index rather than to the pseudo-spin.
At the same time, the intraband polarization functions, $\Pi_{-,-}(q,\,\omega+i0^+)$ and $\Pi_{+,+}(q,\omega+i0^+)$,
vanish due to time-reversal symmetry. This, in turn, results
in the plasmon dispersion in Eq.\,\eqref{EQ:2_7} with electron
Fermi velocity independent of the electron density.
In other words, we have $v_F = \texttt{const}$ as in the
case of 2DTI. Except for the spin factor of $2$, the polarization
function in Eq.\,\eqref{EQ:2_9} is identical to the two-component
HL in 2DTI (i.e., both electron spin states are excited).
From Eqs.\,\eqref{EQ:2_6}, \eqref{EQ:2_7} and Fig.\,\ref{FIG:3},
one can easily understand that the energies of the collective
excitations for the two and single component HLs are well separated
in spectrum, but they share the same particle-hole excitation region.
\medskip

For graphene, the plasmon excitations in doped semiconductor
armchair nanoribbons are similar to the conventional 1DEG plasmon
dispersion. The localization of electron wave functions along two
ribbon edges can be obtained in zigzag nanoribbons. However,
the cone-like band structure disappears for this case.
Moreover, the particle-hole excitations cover a broad region,
instead of a narrow line, and the plasmon dispersion which falls
into this broad region becomes Landau damped.

\section{\label{SEC:4} Single component HL response as a spin-density wave.}

The above paragraph was arguing on strong plasmon branch separation in single and two component electron HL.
However, single component HL is somehow an artificial concept since the topological states always appear as Kramers pair.
Below we shall demonstrate that the density-density plasmon excitations of single component HL are actually
equivalent to the spin-density plasmons in two component HL. As a by-product, we shall also consider spin-spin waves and compare them to those in a conventional 2DEG.
\par
\begin{figure}[]
\centering
\includegraphics[width=0.5\textwidth]{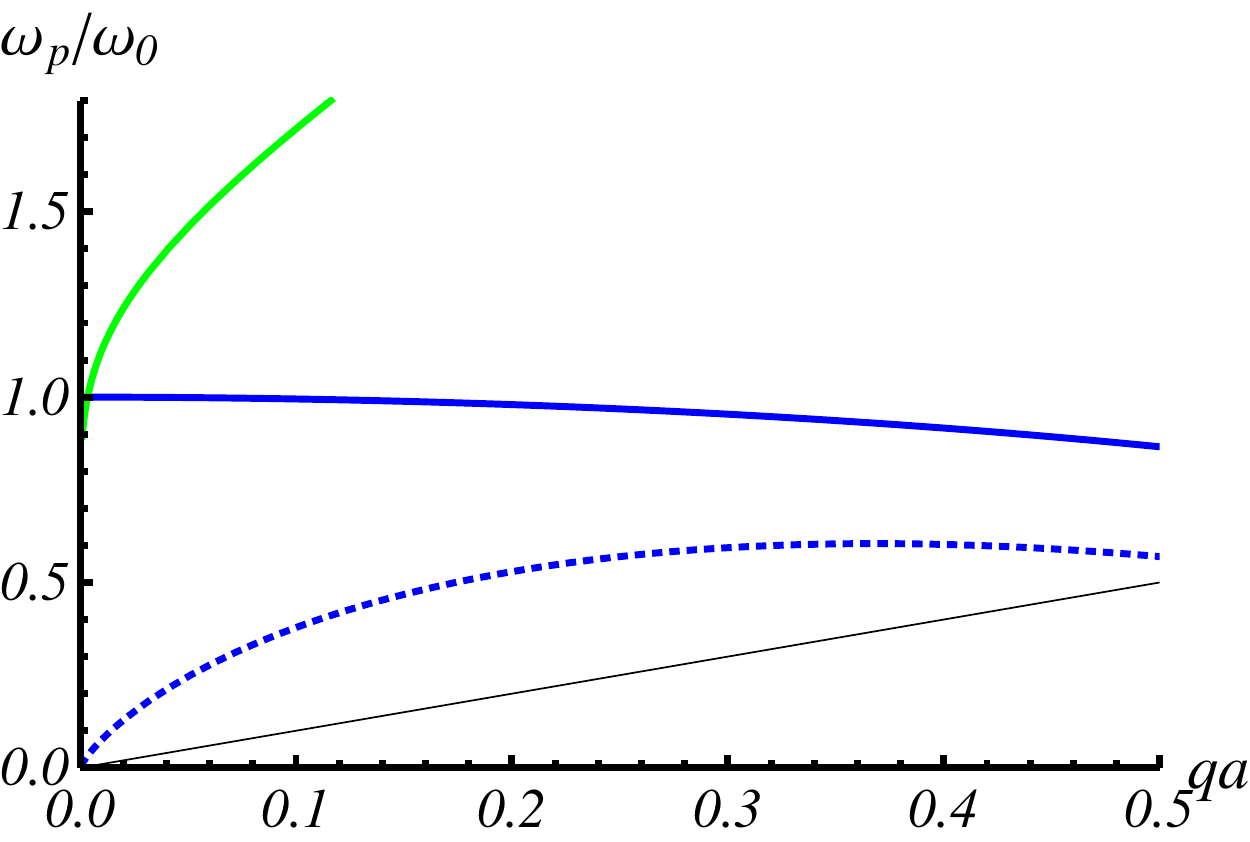}
\caption{\label{FIG:10}
Spin-Spin wave (Eq.\,\eqref{EQ:3_8}) and the plasmons (Eq.\,\eqref{EQ:3_6}) in TI. Thick blue curve correspond to spin-wave with large critical cut-off $1/(2A) V_q \texttt{Log}(2 A k_c) \gg 1$.
Thick green curve is the spin-wave with the cut-off in the linear regime of the dispersion $2 A k_c  =1$.
Dotted curve is the plasmon mode while thin black line is the particle-hole excitations.
}
\end{figure}

Now let us provide theoretical description of the spin-spin and spin-density waves.
For that purpose, we shall need the inter-spin polarizations ($\Pi_{\pm,\mp}$) as well as intra-spin polarizations ($\Pi_{\pm,\pm}$).
The intra-spin polarizations are given in Eq.\,\eqref{EQ:2_9}.
The interspin polarizations are also connected by $\Pi_{+,-}(q,\omega) = \Pi_{-,+}(q,\omega)$.
Here,
\begin{gather}
\label{EQ:3_1}
\Pi_{-,+}(q,\omega)=\int \limits_{-k_c}^{k_c} dk \frac{\theta(E_{-}(k+q))-\theta(E_{+}(k))}{E_{+}(k) - E_{-}(k+q)+\omega}\\
\notag
= - \frac{1}{2 A} \left[{ \texttt{Log}\left({2 A k_c + A q + \omega}\right) + \texttt{Log}\left({-2 A k_c + A q + \omega}\right)}\right]\ ,
\end{gather}
where we have introduced a cut-off wave vector $k_c>q$ due to logarithmic divergence of the above integral when the chemical potential is set to zero.
In the long wave approximation $( Aq + \omega)/2 A k_c \ll 1$, we can simplify the inter-spin polarization as:
\begin{gather}
\label{EQ:3_2}
\Pi_{-,+}(q,\omega)\\
\notag
=
 - \frac{1}{2 A} \left[{ i \pi  + \texttt{Log}\left({2 A k_c}\right) - \frac{1}{4 A^2 k^2_c}\left({\omega + Aq}\right)^2}\right]\ .
\end{gather}
Therefore, two-component helical liquid (Kramers pair) obeys the same response function as multi-level electron plasma\,\cite{Kushwaha2006}. The generalized nonlocal, dynamic dielectric-function matrix connects external $\hat{V}_{ext}$ and induced $\hat{V}_{ind}$ perturbation potential matrix elements:
\begin{gather}
\label{EQ:3_3}
\langle{\nu} \vert {V_{ext}} \vert{\nu'}\rangle =
\sum \limits _{\mu,\mu'} \epsilon^{\rho,\rho}_{\mu \mu'; \nu \nu'} \langle{\mu} \vert {V_{ind}} \vert{\mu'}\rangle\ , \\
\label{EQ:3_2}
 \epsilon^{\rho,\rho}_{\mu \mu'; \nu \nu'} =
\delta_{\mu,\nu'} \delta_{\mu',\nu} - V_q S^{\rho,\rho}_{\mu \mu'; \nu \nu'} \Pi_{\mu,\mu'} (q,\omega)\ .
\end{gather}
Here, the superscript $\rho,\rho$ indicates density-density response; $V_q=V(q)$ in Eq.\,\eqref{EQ:2_4}; the sates $\vert{\mu}\rangle = \vert{\pm}\rangle$ indicate the spin component of the Kramers pair (eigenfunctions of $\hat{\sigma}_z$ Pauli matrix).
The spin overlap function is given by:
\begin{equation}
\label{EQ:3_4}
S^{\rho,\rho}_{\mu \mu'; \nu \nu'} =
\langle{\mu}\vert \sigma_{0} \vert{\mu'} \rangle \langle{\nu} \vert{\sigma_0} \vert{\nu'} \rangle
\end{equation}
with $\sigma_0$ being an identity operator.
Explicitly, Eq.\,\eqref{EQ:3_2} can be written as:
\begin{gather}
\label{EQ:3_5}
 \epsilon^{\rho,\rho} =
 \left({
 \begin{matrix}
 1-V_q \Pi_{+,+} & 0 & 0 &-V_q \Pi_{+,+}\\
 0 & 1 & 0 & 0\\
 0 & 0 & 1 & 0\\
 -V_q \Pi_{-,-} & 0 & 0 & 1-V_q \Pi_{-,-}
 \end{matrix}
 }\right)\ ,
\end{gather}
whose inverse is:
\begin{gather}
\label{EQ:3_6}
\left({\epsilon^{\rho,\rho}}\right)^{-1} \sim
\frac{1}{1-V_q \left({\Pi_{+,+}+\Pi_{-,-}}\right)}\ .
\end{gather}
In order to consider spin-spin response, we shall introduce the spin-lowering $\hat\sigma_{\downarrow}$ and spin-raising $\hat\sigma_{\uparrow}$ operators by their action on the spin states:
\begin{gather*}
\begin{matrix}
\sigma_{\downarrow} \vert{+}\rangle = \vert{-}\rangle; &
\sigma_{\downarrow} \vert{-}\rangle = \vert{0}\rangle\ ,\\
\sigma_{\uparrow} \vert{+}\rangle = \vert{0}\rangle; &
\sigma_{\uparrow} \vert{-}\rangle = \vert{+}\rangle\ ,
\end{matrix}
\\
\langle{0}\vert{\pm}\rangle =0\ .
\end{gather*}
The spin-spin generalized functions are given by Eq.\,\eqref{EQ:3_2} with
\begin{gather}
\label{EQ:3_7}
S^{\uparrow,\downarrow}_{\mu \mu'; \nu \nu'} =
\langle{\mu}\vert \sigma_{\uparrow} \vert{\mu'} \rangle \langle{\nu} \vert{\sigma_{\downarrow}} \vert{\nu'} \rangle\ ,\\
\notag
S^{\downarrow,\uparrow}_{\mu \mu'; \nu \nu'} =
\langle{\mu}\vert \sigma_{\downarrow} \vert{\mu'} \rangle \langle{\nu} \vert{\sigma_{\uparrow}} \vert{\nu'} \rangle\ .
\end{gather}
The explicit form of the dielectric function is:
\begin{gather}
\epsilon^{\uparrow,\downarrow} = \texttt{diag}\left[{1,1-V_q \Pi_{+,-},1,1}\right]\ ,\\
\notag
\epsilon^{\uparrow,\downarrow} = \texttt{diag}\left[{1,1,1-V_q \Pi_{-,+},1}\right]\ ,
\end{gather}
and the spin-spin waves are given by real zeros of :
\begin{gather}
\label{EQ:3_8}
\left({\epsilon^{\uparrow,\downarrow}+\epsilon^{\downarrow,\uparrow}}\right)^{-1} \sim
\frac{1}{1-V_q \left({\Pi_{+,-}+\Pi_{-,+}}\right)}\ .
\end{gather}
Dispersion curves of the spin-spin waves are shown in Fig.\,\ref{FIG:10}
Note that in the limit of large cut-off wave vector  $1/(2A) V_q \texttt{Log}(2 A k_c) \gg 1$, the spin-wave is analogues to
$\Omega_-$ inter-spin branch in case of Rashba spin-orbit\,\cite{Kushwaha2006}.
The plasmon branch is similar to $\Omega_0$ of intra-spin plasmon. $\Omega_-$ inter-spin branch has no analogue in TI case.
The reason to that is four possible inter-spin transitions in the Rashba split 2DEG as opposite to two of those in TI.
\par
The spin-density response is defined by Lozovik\,\cite{efimkin2011spin} as Eq.\,\eqref{EQ:3_2} with the spin overlap given by:
\begin{gather}
\label{EQ:3_7}
S^{\rho,\bar{\sigma}}_{\mu \mu'; \nu \nu'} =
\langle{\mu}\vert \sigma_{0} \vert{\mu'} \rangle \langle{\nu} \vert{\bar{\sigma}} \vert{\nu'} \rangle\ ,
\end{gather}
 where $\bar{\sigma} = \left[{\sigma_x=\sigma{\uparrow}+\sigma{\downarrow},\sigma_y=-i\left({\sigma{\uparrow}-\sigma{\downarrow}}\right),\sigma_z}\right]$.
One can show that
\begin{gather}
\label{EQ:3_8}
\epsilon^{z,z}=\epsilon^{\rho,\rho}\ ,\\
\notag
(\epsilon^{x,x})^{-1}\sim \frac{1}{1-V_q \left({\Pi_{+,-}+\Pi_{-,+}}\right)}\ ,\\
\notag
(\epsilon^{y,y})^{-1}\sim \frac{1}{1-V_q \left({\Pi_{+,-}+\Pi_{-,+}}\right)}\ .
\end{gather}
Due to the fact that in 3DTI the spin wave function component depends on wave vector, we are left with only transverse $\langle \sigma^{z} \rangle$ component of the response survive.
In our case, those actually provide the spin-spin waves.
Those spins may be coupled to the density via $\epsilon^{\rho,x},\epsilon^{\rho,y}$. However, the inverse of those matrices does \emph{not} provide any resonances, and then, such coupling may be neglected.
On the other hand, we find the rest inverse as:
\begin{gather}
\label{EQ:3_9}
(\epsilon^{\rho,z})^{-1}\sim \frac{1}{1-V_q \left({\Pi_{-,-}-\Pi_{+,+}}\right)}\ ,\\
\label{EQ:3_10}
(\epsilon^{\rho,-z})^{-1}\sim \frac{1}{1-V_q \left({\Pi_{+,+}-\Pi_{-,-}}\right)}\ .
\end{gather}
In the above two equations, the time reversal symmetry is broken, similar (but not identical)  to the case of single component HL discussed in the previous section.
There are four spin-plasmon modes:
\begin{gather}
\pi \omega_p^{\rho,\pm z} = \mp q v_q \pm \sqrt{(\pi A q)^2 + (q V_q)^2}\ .\\
\end{gather}
Apart from a factor of two, in the long wave approximation we shall obtain the mode given by the single component HL:
\begin{equation}
\omega^{\rho,-z}_p = \frac{2}{\pi} q V_q = -2 \omega_0 q \texttt{ln}(q)\ .
\end{equation}
Due to Eq.\,\eqref{EQ:3_8} the combined spin-density response can be written via the spin-density overlap as:
\begin{gather}
\label{EQ:3_11}
S^{\rho,\rho\pm\sigma_z}_{\mu \mu'; \nu \nu'} =
\langle{\mu}\vert \sigma_{0} \vert{\mu'} \rangle \langle{\nu} \vert{\sigma_0 \pm \sigma_z} \vert{\nu'} \rangle\ ,
\end{gather}
so that we recover the single-component HL response:
\begin{equation}
(\epsilon^{\rho,\rho \pm z})^{-1}\sim \frac{1}{1-V_q \Pi_{\mp,\mp}}\ .
\end{equation}
To conclude this paragraph we have demonstrated that the spin-density response mimics that of single component helical liquid.
Also we found similar behavior of spin-spin waves in TI to those provided by Rashba spin-orbit split in a conventional 2DEG.

\section{Concluding Remarks}
\label{SEC:5}

The dispersion of intraband plasmon excitations in a semi-infinite
inverted $\texttt{HgTe/CdTe}$ quantum well has been derived within
the random-phase approximation based on a calculated edge-localized
topological state of electrons in a single-component helical liquid using the BHZ model.
Under the perturbation from a linearly-polarized incident light,
the unique properties in the collective excitation of these edge-bound electrons with a broken time-reversal symmetry has been explored.
Our calculations predict the plasmon dispersion  $\omega_p(q)\sim-\omega_0q\,\texttt{ln} (qa)$ for such a single-component helical state in the long-wavelength limit, in sharp contrast
with $\omega_p(q)\sim-\omega_0 \sqrt{- \texttt{ln} (q W)}$ found for a one-dimensional electron gas in a quantum-wire system.
Moreover, $\omega_0$ in our plasmon dispersion is independent of the linear electron density, similar to the case for a metallic armchair graphene nanoribbon.
On the other hand, the plasmon dispersion of the two-component helical liquid is found to be identical to that of a armchair graphene nanoribbon
except for the spin perfector and a characteristic-width scaling of the wave number.
The particle-hole excitation region shrinks into a straight line in our system, in comparison with a wide region for a conventional one-dimensional electron gas.
The plasmon energy of the single-component helical state is well separated from that of the two-component helical state but they share the common particle-hole excitation region in the excitation spectrum.
The density-density plasmon excitations of single component HL are equivalent to the spin-density plasmons in two component HL.

\acknowledgments

This research was supported by the contract \# FA9453-11-01-0263 of AFRL. DH would like
to thank the Air Force Office of Scientific Research (AFOSR) for its support.

\bibliography{biblio}

\end{document}